\documentclass[lettersize,journal]{IEEEtran}
\usepackage{amsmath,amsfonts}
\usepackage{algorithmic}
\usepackage{algorithm}
\usepackage{array}
\usepackage[caption=false,font=normalsize,labelfont=sf,textfont=sf]{subfig}
\usepackage{textcomp}
\usepackage{stfloats}
\usepackage{url}
\usepackage{verbatim}
\usepackage{graphicx}
\usepackage{multirow}
\usepackage{booktabs}
\usepackage{bbding}
\usepackage{color}
\usepackage{cite}
\usepackage{float}
\usepackage[backref]{hyperref}
\begin{document}

\title{Real-time High-resolution View Synthesis of Complex Scenes with Explicit 3D Visibility Reasoning}

\author{Tiansong Zhou, Yebin Liu, Xuangeng Chu, Chengkun Cao, Changyin Zhou, Fei Yu, Yu Li
\thanks{Corresponding author: Yu Li.}
\thanks{Tiansong Zhou, Xuangeng Chu, Chengkun Cao, Yu Li are with International Digital Economy Academy. Email: tszhou@foxmail.com; xuangeng.chu@mi.t.u-tokyo.ac.jp; 
chengkun$\_$cao@outlook.com; 
liyu@idea.edu.cn}
\thanks{Yebin Liu is with Tsinghua University. Email: liuyebin@mail.tsinghua.edu.cn.}
\thanks{Changyin Zhou, Fei Yu are with Vistring Inc. Email:changyin.zhou@gmail.com; yufei.flyingfish@gmail.com}

}

\markboth{Journal of \LaTeX\ Class Files,~Vol.~14, No.~8, August~2021}%
{Shell \MakeLowercase{\textit{et al.}}: A Sample Article Using IEEEtran.cls for IEEE Journals}


\maketitle
\begin{figure*}[h]
    \centering
    \includegraphics[width=0.95\linewidth]
    {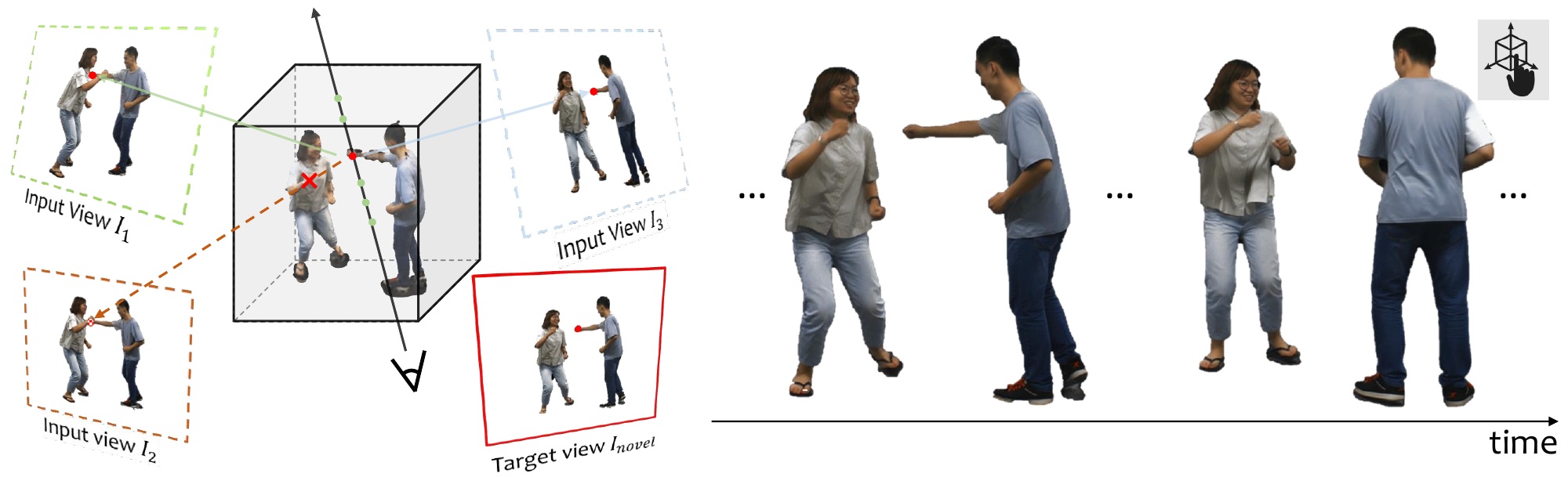}
    \caption{
        \textbf{Our method achieves real-time for rendering high-resolution dynamic scenes with high visual quality,  enabling users to seamlessly transition to desired perspectives at any time.} At the heart of our method is explicit 3D visibility reasoning, which efficiently estimates the visibility of the sampled 3D points to the input views, helping us to address the occlusion problems arising from the sparsity of input views and the complexity of captured scenes.
    }
    \label{fig:teaser}
\end{figure*}

\begin{abstract}
Rendering photo-realistic novel-view images of complex scenes has been a long-standing challenge in computer graphics. 
In recent years, great research progress has been made on enhancing rendering quality and accelerating rendering speed in the realm of view synthesis.
However, when rendering complex dynamic scenes with sparse views, the rendering quality remains limited due to occlusion problems.
Besides, for rendering high-resolution images on dynamic scenes, the rendering speed is still far from real-time.
In this work, we propose a generalizable view synthesis method that can render high-resolution novel-view images of complex static and dynamic scenes in real-time from sparse views.  
To address the occlusion problems arising from the sparsity of input views and the complexity of captured scenes, we introduce an explicit 3D visibility reasoning approach that can efficiently estimate the visibility of sampled 3D points to the input views.
The proposed visibility reasoning approach is fully differentiable and can gracefully fit inside the volume rendering pipeline, allowing us to train our networks with only multi-view images as supervision while refining geometry and texture simultaneously. 
Besides, each module in our pipeline is carefully designed to bypass the time-consuming MLP querying process and enhance the rendering quality of high-resolution images, enabling us to  render high-resolution novel-view images in real-time.
Experimental results show that our method outperforms previous view synthesis methods in both rendering quality and speed, particularly when dealing with complex dynamic scenes with sparse views.
\end{abstract}


\section{Introduction}
Novel-view rendering, allowing users to navigate through rendered scenes as in the real world, has garnered significant research attention for many years. For practical applications, novel-view rendering technologies must fulfill two key requirements. Firstly, the rendered results should exhibit such realism that users cannot distinguish them from reality. Secondly, the rendering speed should be real-time or interactive, enabling quick transitions to desired perspectives.

Recently, Neural Radiance Fields (NeRF) \cite{nerf} is the most promising method for novel-view rendering, which shows impressive photo-realistic rendering results by using multi-layer perceptrons (MLPs) to represent 3D scenes. 
However, vanilla NeRF needs to train a separate network for each scene.
When dealing with dynamic scenes, it needs to train a network for each frame individually, which is impractical.
Some works \cite{dnerf, hyperreel, gao2021dynamic} seek to extend the NeRF framework to dynamic scenes by using the timestamp as an additional input, in which they learn a network for each sequence separately.
However, the representation capability of neural networks is always restricted, so these works are only able to handle a limited number of frames. 
When dealing with a long sequence of high-resolution complex scenes, which needs a large network capacity for representation, the networks might be overwhelmed and result in rendering failures.

These difficulties have led to the generalizable NeRF methods \cite{pixelnerf, ibrnet, enerf, neuralrays}. 
This line of works is able to generalize across scenes or frames that are unseen in the training data. 
When dealing with dynamic scenes, they simply treat each frame as an independent scene and render novel views. 
With the ability to generalize across frames, they are able to handle sequences with arbitrary length. 
To get generalization ability, most generalizable NeRF methods follow a typical high-level pipeline, involving encoding input views to feature space, projecting sampled 3D points to image space to grab features, and aggregating the grabbed multi-view features. 
The heart of this pipeline is how to weigh each view's feature in the stage of multi-view aggregation, which raises the classical occlusion problem in 
the view synthesis task. Ideally, if a sampled 3D point is occluded by surfaces to one view, the weight of the corresponding view should be lower, and vice versa.

To solve the occlusion problems the in generalizable NeRF pipeline, IBRNet \cite{ibrnet} and ENeRF \cite{enerf} use an MLP to predict the visibility of the sampled 3D points to input views based on the feature consistency between views.
However, when the input views are relatively sparse and the scenes are highly complex, the grabbed multi-view features might be highly inconsistent, making the predicted visibility to be inaccurate and resulting in rendering artifacts. 
To perform occlusion-aware radiance fields construction, NeuRay \cite{neuralrays} learns a visibility feature map for each input view and uses an MLP to predict the distribution of surfaces location from the visibility features.
This visibility reasoning process is actually performing per-view reconstruction, which can not guarantee the global consistency of the reconstruction between views.
In our experiments, we found that when dealing with high-resolution complex scenes, the per-view reconstruction seems to be unreliable, exacerbating the global inconsistency problems. 
The global inconsistency will cause inaccurate visibility estimation, decreasing the rendering quality of NeuRay.

In this paper, we present a view synthesis method to render high-resolution novel-view images of complex static and dynamic scenes in real-time from sparse views.
Similar to \cite{mvsnerf, enerf, neuralrays}, our system is a generalizable rendering method that is based on the volume rendering technique.
To solve the occlusion problems arising from the sparsity of input views and the complexity of captured scenes, we propose a novel technique to perform visibility reasoning.
We first learn a density volume in the novel view, then re-sampling the density volume to input views, and calculate a visibility volume for each view to get the visibility information.
Different from the above-mentioned generalizable rendering works \cite{ibrnet, enerf, neuralrays} that use implicit methods to perform visibility reasoning, where the visibility information is predicted by implicit MLPs, our visibility reasoning is an explicit 3D approach as the visibility is calculated through the explicitly constructed 3D volumes.
Such an explicitly 3D visibility reasoning method is globally consistent as all the calculated visibility volumes derive from the same novel view's density volume.
This consistency helps us to estimate accurate visibility, enhancing our rendering quality and stability.
Besides, though the proposed explicit 3D visibility reasoning does not contain any network, it is differentiable and can gracefully fit inside the volume rendering technique, enabling us to optimize the geometry and texture simultaneously during the training time.

\begin{table}[]
\caption{\textbf{Key aspects of our method relative to previous works.} 
}
\centering
\begin{tabular}{lccc}

\toprule
           & Generalizable?                    & Dynamic?                          & Real-time?       \\ \hline
MVSNeRF    & \textcolor{green}{\CheckmarkBold} & Arbitrary & \XSolidBrush     \\
NeuRay     & \textcolor{green}{\CheckmarkBold} & Arbitrary & \XSolidBrush     \\
ENeRF      & \textcolor{green}{\CheckmarkBold} & Arbitrary & Low-Res    \\
InstantNGP & \XSolidBrush                      & \XSolidBrush                      & High-Res \\
D-NeRF     & \XSolidBrush                      & Limited & \XSolidBrush     \\
Ours       & \textcolor{green}{\CheckmarkBold} & Arbitrary & High-Res   \\ \bottomrule
\end{tabular}

\vspace{0.1cm}
\raggedright
"Arbitrary" and "Limited" indicate the method is able to handle dynamic scenes with arbitrary and limited frame lengths respectively.
"Low-Res" and "High-Res" indicate rendering images at low and high resolution respectively.
\vspace{-0.2 cm}
\label{tab:aspects}
\end{table}

Apart from the rendering quality, another unsolved challenge is the speed of rendering dynamic high-resolution complex scenes.
Though extensive effects \cite{fastnerf, snerg, mobilenerf, instantngp} have been made on accelerating vanilla NeRF, they are only applicable to static scenes.
Like vanilla NeRF, most generalizable methods \cite{mvsnerf, neuralrays} need to query the MLP millions of times, making their rendering speed extremely slow.
A more similar work to our method is ENeRF \cite{enerf}, which uses a depth-guided strategy to reduce the sampled points to accelerate generalizable NeRF methods, achieving real-time when rendering images at a resolution of 512$\times$512.
However, ENeRF still needs to query MLP for each sampled point, meaning that the rendering speed is inversely linear to the number of rendered pixels. 
When rendering high-resolution images, where there are too many pixels needed to process, the rendering speed of ENeRF is still far from real-time.
In our method, we perform ray integration in the feature space, generating a feature map for the target view and employing a 2D CNN to render the final color images. 
Integrating rays on feature space allows us to bypass the computationally expensive MLP querying process, leading to accelerated rendering speeds.
Additionally, the 2D CNN serves as a texture prior for the rendered scenes, enhancing rendering quality and stability, particularly for high-resolution images.
By integrating rays on feature space to accelerate rendering speeds and the 2D CNN to generate high-resolution images, our method achieves \textit{high-resolution} novel-view rendering in \textit{real-time}.
Key aspects of our method relative to previous works are summarized in Table~\ref{tab:aspects}.

We conduct extensive experiments on both static and dynamic scenes. For static scenes, we evaluate our method on the DTU \cite{dtu}, NeRF synthetic \cite{nerf} and Real Forward-facing datasets \cite{llff}. The results demonstrate that our pre-trained network can be efficiently fine-tuned for new scenes, yielding competitive performance under per-scene fine-tuning settings. Regarding dynamic scenes, our method effectively handles complex scenarios with severe occlusions, producing high-quality spatial-temporal coherent rendering results.
In terms of rendering speed, our method achieves real-time performance in rendering high-definition images at a resolution of 1280$\times$720. Moreover, it can also deliver interactive frame rates while rendering full high-definition images at a resolution of 1920$\times$1080.

Our contributions can be summarized as follows:
\begin{itemize}
\item We propose a view synthesis method capable of real-time rendering of high-resolution novel-view images from sparse view inputs.
\item Our method employs explicit 3D visibility reasoning as the core technique to address occlusion problems arising from the sparsity of input view and the complexity of captured scenes. This technique is fully differentiable, which can be jointly learned with the volume rendering techniques. 
\item Experimental results demonstrate significant improvements in both rendering quality and speed compared to previous approaches, particularly for complex dynamic scenes.
\end{itemize}

\section{Related Work}
\subsection{NeRF Works}
Recently, Neural radiance fields (NeRF) \cite{nerf} is the most promising view synthesis method, which achieves state-of-the-art rendering quality and attracts a lot of attention. 
NeRF uses MLPs to represent scenes by mapping 3D points and view directions to density and color values.
As a compact and lightweight representation, NeRF has inspired substantial follow-up works. These works can be divided into three categories:

\textbf{Scene-specific NeRF.}
Many works \cite{4knerf, mipnerf, mipnerf360, dsnerf, densedepth, refnerf} focus on improving the performance of vanilla NeRF.
4K-NeRF \cite{4knerf} tries to boost the representation capacity of vanilla NeRF by exploring correlations between rays to extend the NeRF-based paradigm to 4K resolution.
DS-NeRF \cite{dsnerf} adopt the reconstructed sparse point clouds from COLMAP \cite{pixelwise, schoenberger2016sfm} to add an additional depth loss to improve the reconstruction and rendering quality of NeRF.
Mip-NeRF \cite{mipnerf} and Mip-NeRF 360 \cite{mipnerf360} integrate the mipmap technique into the NeRF framework to achieve anti-aliasing.
This line of works is only applicable to render static scenes as they need to train an independent network for each scene.

\begin{figure*}[t]
    \centering
    \includegraphics[width=0.95\linewidth]
    {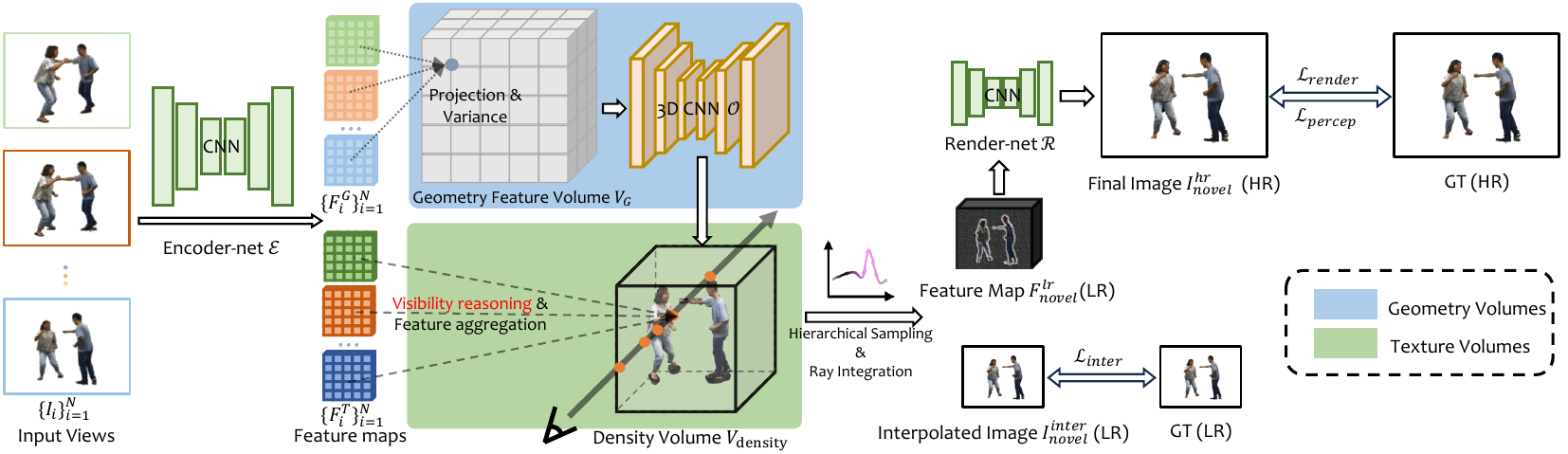}
    \vspace{-0.1cm}
    \caption{
        \textbf{The overall pipeline of our system.}
        Our pipeline firstly uses an encoder-net $\mathcal{E}$ to extract geometry and texture feature maps from the input images. In the geometry volumes branch, we construct a 3D feature volume on the novel view's camera frustum based on the extracted geometry feature maps. Then, we use a 3D CNN $\mathcal{O}$ to regress a density volume of the novel view.
        In the texture volumes branch, we hierarchically sample 3D points in each marching ray of the novel view.
        Then, we project each sampled 3D point to input views and grab features on the texture feature maps.
        To aggregate the grabbed multi-view features, we use explicit 3D visibility reasoning to get the weight of each view.
        After aggregating multi-view features and performing ray integration for all the rays, we get a low-resolution feature map $F_{novel}^{lr}$. 
        Apart from the feature map, we also interpolate a low-resolution color image $I_{novel}^{inter}$ as an additional supervision signal to make our pipeline more geometrically interpretable.
        Finally, we use a render-net $\mathcal{R}$ to render the high-resolution color image $I_{novel}^{hr}$ from the low-resolution feature map $F_{novel}^{lr}$, in which we procedurally up-sample the $F_{novel}^{lr}$ to high resolution in the render-net $\mathcal{R}$. 
        Our method is fully differentiable and can be trained with only sparse multi-view images as supervision.
    }
    \label{fig:pipeline}
\end{figure*}

\textbf{Sequence-specific NeRF.}
Some works \cite{dnerf, park2021nerfies, hyperreel} seek to extend NeRF to a temporal domain to process dynamic scenes.
D-NeRF \cite{dnerf} represents the input with a continuous 6D function, including a 3D location $(x, y, z)$, 2D viewing direction $(\theta, \phi)$, and time component $t$. 
It maps each frame to a canonical space and performs novel-view rendering in the canonical space.
Nerfies \cite{park2021nerfies} anchors a latent code to each frame to handle non-rigidly deforming objects by optimizing a deformation field per observation.
HyperReel \cite{hyperreel} proposes a memory-efficient dynamic volume representation by exploiting the spatial-temporal redundancy of a dynamic scene.
To achieve real-time rendering, it additionally learns a ray-conditioned sample prediction network that predicts sparse point samples for volume rendering. 
These works make great success in handling dynamic scenes and some of them even achieve real-time rendering while keeping a low memory footprint. 
However, the capacity of a neural network is always restricted no matter how deep the network is, so these sequence-specific works are only able to handle a limited number of frames.
When facing a long sequence of highly complex dynamic scenes, the networks might be overwhelmed and result in rendering failures.

\textbf{Generalizable NeRF.}
Some works \cite{pixelnerf, mvsnerf, ibrnet, enerf, neuralrays} extend NeRF to generalizable rendering methods that are able to generalize across scenes and frames.
PixelNeRF \cite{pixelnerf} conditions the MLP with encoded features of input views to get generalization ability.
MVSNeRF \cite{mvsnerf} combines NeRF with learning-based multi-view stereo (MVS) methods to achieve highly efficient radiance fields construction.
ENeRF \cite{enerf} seeks to accelerate generalizable NeRF methods to achieve real-time rendering of dynamic scenes.
NeuRay \cite{neuralrays} proposes an occlusion-aware radiance fields construction method.
Our method also belongs to this line of works, which is able to handle both static and dynamic scenes.

\begin{figure*}[htbp]
    \centering
    \includegraphics[width=0.95\linewidth]
    {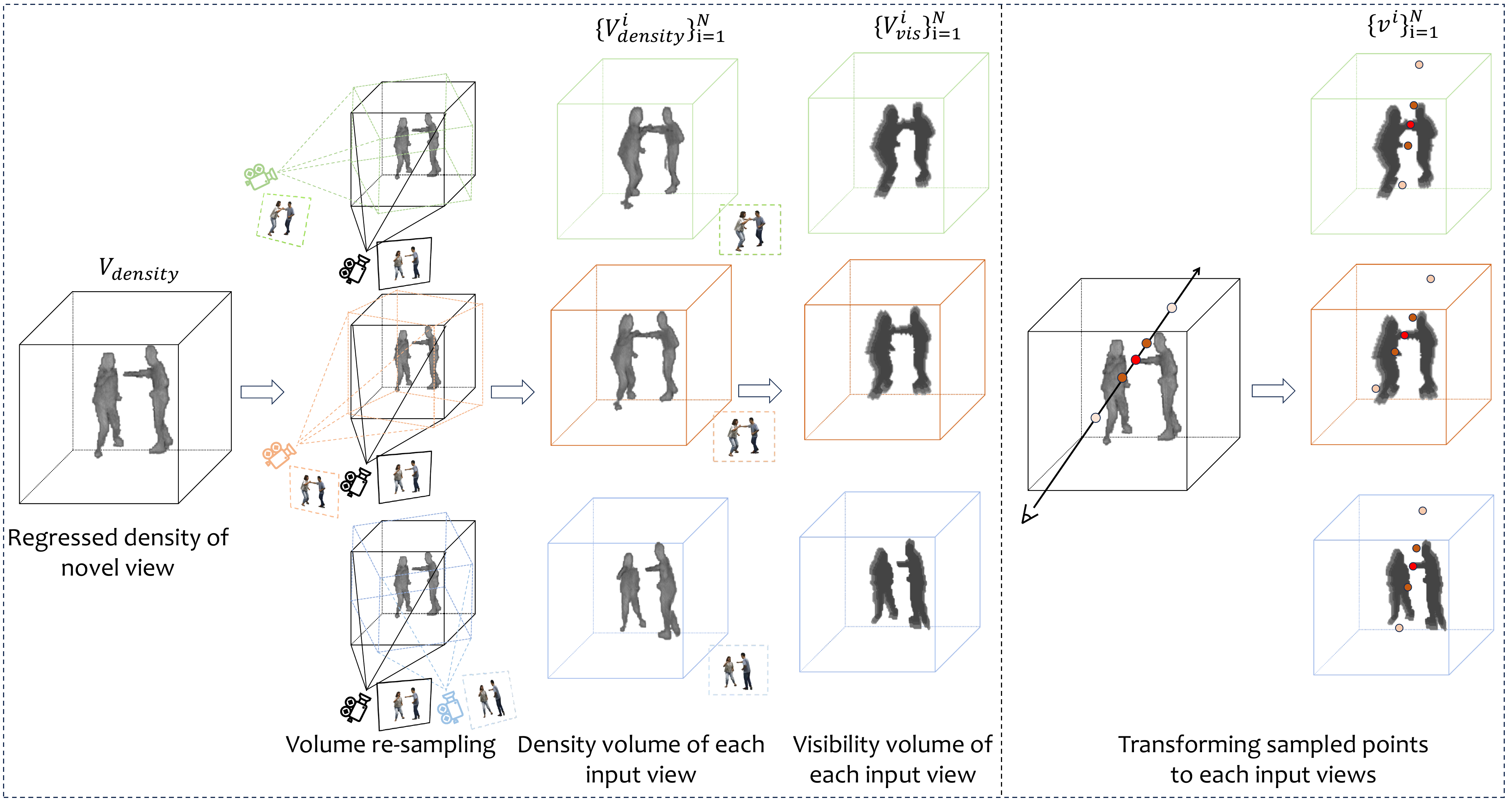}
    \caption{\textbf{Our explicit 3D visibility reasoning.} Based on the regressed density volume on the novel view (the first column), we build a volume in each input view's camera frustum and use the constructed volume to re-sample the novel view's density volume (the second column), getting the density volumes of input views (the third column). Then, based on the re-sampled density volumes, we calculate the visibility volumes (the fourth column) using Equation~\ref{eq:alpha} and Equation ~\ref{eq:visibility}. 
    Finally, for each hierarchically sampled ray in the novel view (the fifth column), we transform the sampled 3D points to input views' visibility volumes and get visibility weights by tri-linearly interpolating the visibility volumes (the sixth column). }
    \label{fig:visibility}
\end{figure*}

\subsection{NeRF Acceleration}
A main drawback of vanilla NeRF is that it needs to query MLP millions of times to render an image, which is time-consuming and makes its rendering speed extremely slow. 
To make NeRF more suitable for practical usages, many works \cite{snerg, fastnerf, nsvf, kilonerf, derf, instantngp, neuralsceneflow, du2021nerflow} seek to accelerate NeRF to achieve fast rendering speed. 
NSVF \cite{nsvf} defines a set of voxel-bounded implicit fields in 3D space, enabling it to use classical acceleration strategies like empty space skipping and early ray termination. 
KiloNeRF \cite{kilonerf} uses thousands of tiny MLPs to replace a large MLP to represent a complex scene and each tiny MLP only represents a small part of the scene. The tiny MLPs help to reduce computation consumption and decrease rendering times. 
DeRF \cite{derf} decomposes a scene into a few small parts and uses a small MLP to represent each part. Similar to KiloNeRF, the small MLPs help it to reduce the rendering time. 
SNeRG \cite{snerg} trains a NeRF for a scene first, then "bakes" it to a sparse neural radiance grid for acceleration.
InstantNGP \cite{instantngp} proposes a multi-resolution hash map to represent a 3D embedding.
It decodes features fetched from the 3D embedding with a small MLP, achieving amazing super-fast training and real-time rendering.
Unfortunately, all the acceleration strategies of the above-mentioned works are restricted to static scenes.

As a generalizable NeRF rendering method, ENeRF \cite{enerf} uses a depth-guided sampling strategy to reduce the number of sampled points to accelerate the generalizable NeRF pipeline, achieving real-time when rendering images at a resolution of $512\times512$.
However, when dealing with high-resolution images, its rendering speed is still far from real-time as there are many pixels to process.
In our method, we perform ray integration on feature space to bypass the time-consuming MLP querying stage, then use a 2D CNN to render high-resolution images in real-time.

\subsection{Visibility Reasoning}
Occlusion is a long-standing problem in the field of image-based modeling and rendering. 
The importance of generating accurate results at occlusion edges has been identified in many early IBR works \cite{viewbased, parallax, highquality}.
In classic IBR pipelines, solving occlusion problems usually lies in performing visibility reasoning and improving the blending weights of the warped images or features in the stage of multi-view fusion.
Early works use soft visibility \cite{viewbased, parallax} or alpha matting \cite{highquality} to improve the blending weights.
More recently, Soft3D \cite{soft3d} proposes a sophisticated blending approach using a soft estimation of visibility.
In the past decade, deep learning has shown very impressive and promising results in many tasks of computer vision and graphics and it has been applied to view synthesis problems.
Unfortunately, the visibility reasoning in the early IBR methods is indifferetiable, making them unable to be directly integrated into learning-based pipelines.
To this end, deep blending
\cite{deepblending} utilizes a convolution neural network (CNN) to predict a visibility map for each warped input image and uses the visibility maps as blending weights.
FVS \cite{fvs} uses a CNN to estimate the blending weight maps for the warped feature maps of input views.

In recent years, NeRF \cite{nerf} and volume rendering \cite{drebin1988volume} have attracted great research attention as they show very impressive and convincing rendering quality.
Many works \cite{pixelnerf, mvsnerf, ibrnet, enerf, neuralrays} seek to construct generalizable radiance fields, which are able to generalize across scenes or frames.
For the generalizable NeRF works, they usually follow a typical rendering pipeline, including encoding the input images to feature space, projecting sampled 3D points to image space to grab features, and aggregating the grabbed multi-view features.
At the heart of this pipeline is how to get the aggregation weights of multi-view features.
Ideally, the weights should reflect the visibility of the sampled 3D points to the input views.
However, the visibility reasoning in many learning-based methods \cite{deepblending} can not fit inside the volume rendering technique as the learned visibility maps are in 2D space whereas in volume rendering, the visibility reasoning for the sampled 3D points should be operated in 3D space.
IBRNet \cite{ibrnet} and ENeRF \cite{enerf} use an MLP to predict the visibility of each sampled 3D point to input views based on the consistency of the grabbed features.
This method can only achieve high-quality rendering when the input views are relatively dense.
When the input views are sparse, the inconsistency between multi-view features becomes more severe, making the predicted visibility to be inaccurate.
To perform occlusion-aware radiance fields construction, NeuRay \cite{neuralrays} learns a visibility feature map for each input view and uses an MLP to estimate the distribution of surfaces location from the visibility features.
This process is actually performing per-view reconstruction, which can not guarantee global consistency of the reconstruction between views.
The inconsistency problem may cause inaccurate visibility estimation and result in rendering artifacts.
In contrast, the proposed explicit 3D visibility reasoning in our method is fully consistent as all the visibility information derives from the same reconstructed density volume of the novel view, which provides accurate visibility estimation and helps us to improve our rendering quality and stability.

\section{Method}
To address the occlusion problems arising from the sparsity of input view and the complexity of captured scenes, our method focuses on solving visibility estimation. This requires knowledge of surface locations, leading us to divide the volume rendering technique into discretized geometry volumes and continuous texture volumes. In the geometry volumes, we perform initial geometry reconstruction. Building upon the results of this reconstruction, we employ explicit 3D visibility reasoning to address the occlusion problems in the texture volumes. 
In the following, we will introduce the encoding CNN at Section~\ref{sec:encoding}, the geometry volumes at Section~\ref{sec:geometry_volumes}, the texture volumes at Section~\ref{sec:texture_volumes}, followed by the rendering CNN at Section~\ref{sec:rendering}, and our training loss at Section~\ref{sec:training_loss}.

\subsection{Feature Encoding}
\label{sec:encoding}
Given a novel view $C_{novel}$, we first select $N$ nearest neighbor views to the novel view as the input of our system among all the captured $K$ views.
In our experiments, we set $N=3$ for static scenes and $N=4$ for dynamic scenes.
Then, we encode all the input images into a feature space using a 2D encoder-net $\mathcal{E}$ as:
\begin{equation}
    F_i^G, F_i^T = \mathcal{E}(I_i),
    \label{eq:encoder}
\end{equation}
where $I_i$ is the image of input view $i$ and $F_i^G, F_i^T$ are the corresponding encoded geometry feature map and texture feature map, respectively. 
We adopt a shared encoder for both the geometry and texture, and differentiate them using two parallel convolution layers in the final layer. This allows efficient computation of shared representation at early layers while maintaining separate features for the geometry and texture information in the final stage of encoding.

\subsection{Discretized Geometry Volumes}
\label{sec:geometry_volumes}
In the geometry volumes, we first build geometry feature volumes using the plane-sweep algorithm in the novel view, followed by a 3D CNN to map the feature volumes to density volumes.

\subsubsection{Feature volume construction}
Given the camera parameters $C_{novel}$ of the novel view, we construct a discretized volume in the novel view's camera frustum. Specifically, for a voxel $P\in{\mathbb{R}^3}$, we get its position at world coordinate through:
\begin{equation}
    P(u,v,d) = C_{novel}^{-1}(u, v, \zeta(d)),
    \label{eq:construction}
\end{equation}
\begin{equation}
    \zeta(d)= t_n + d * \frac{t_f - t_n}{D},
    \label{eq:cal_depth}
\end{equation}
where $(u, v, d)$ is the voxel coordinate of novel view. $C^{-1}_{novel}$ is the inverse projection of novel view. To perform inverse projection, we use $\zeta$ to calculate the depth value of plane $d$. $t_n, t_f$ are the near and far planes respectively. $D$ is the number of sampled planes. 

Then, we project $P$ to each input view and bilinearly sample the encoded feature maps as:
\begin{equation}
    f_i^G(u,v,d) = F_i^G(C_i(P(u, v, d))),
\end{equation}
where $C_i$ is the forward projection of view $i$. 
We compute the variance of the multi-view features ${\{f_i^G\}}_{i=1}^N$ for multi-view feature fusion, where $N$ is the number of input views.
After getting the feature vectors of all the voxels, we get a geometry feature volume $V_G \in \mathbb{R}^{{H_G}\times{W_G}\times{D_G}\times{C_G}}$, where $C_G=32$ is the number of feature channels and $H_G, W_G$ are hyper-parameters. 
$D_G=D$ is the number of the sampled depth planes and we set it to 96 or 128 in our experiments.

\subsubsection{Density volume regression}
After constructing the geometry feature volume $V_G$, we use a 3D CNN $\mathcal{O}$ to regress the density volume $V_{density}\in\mathbb{R}^{{H_G}\times{W_G}\times{D_G}\times{1}}$ as:
\begin{equation}
    V_{density} = \mathcal{O}(V_G).
    \label{eq:density}
\end{equation} 

\subsection{Continuous Texture Volumes}
\label{sec:texture_volumes}
Based on the reconstructed density volume, we introduce our continuous texture volumes.

\subsubsection{Ray hierarchical sampling}
For each marching ray of the novel view, we first sample along the ray uniformly between the near and far planes, getting $N_{u}$ points. 
Then, we get the density value of each sampled point by tri-linearly sampling the density volume $V_{density}$. 
Based on the density values, we perform hierarchical sampling following vanilla NeRF \cite{nerf} and get $N_{h}$ points. 
Also, for each point $P_s \in N_{h}$, we get its density value $\sigma_s$ by tri-linearly sampling the density volume $V_{density}$.
We set $N_{u} = 128$ or $96$ and $N_{h} = 8$ in all the experiments.

\subsubsection{Explicit 3D visibility reasoning}
For each sampled point $P_{s}\in\mathbb{R}^3$ in $N_{h}$, we project it to each input view to retrieve features and RGB values using bilinear sampling as:
\begin{equation}
    f^T_{i, s}, f^{RGB}_{i, s} = F_i^T(C_i(P_s)), I_i(C_i(P_s)).
\end{equation}
Then, we perform multi-view feature aggregation:
\begin{equation}
    f^T_s, f^{RGB}_s = \frac{1}{W}\sum_{i=1}^N v_i*f^T_{i, s}, \frac{1}{W}\sum_{i=1}^N v_i*f^{RGB}_{i, s},
    \label{eq:aggregation}
\end{equation}
where $v_i$ is the visibility weight of view $i$ and $W=\sum_{i=1}^N v_i$ for normalization. 
In the following, we will introduce how to get visibility weights $\{v_i\}_{i=1}^N$ using explicit 3D  visibility reasoning.

The procedure of explicit 3D visibility reasoning is shown in Figure~\ref{fig:visibility}.
We start by constructing a volume in each input view's camera frustum using Equation~\ref{eq:construction}, in which $C_{novel}$ should be replaced with $C_i$, and we get a volume $V^i\in\mathbb{R}^{{H_G}\times{W_G}\times{D_G}\times3}$ in each input view $i$'s camera frustum. 
Then, for each voxel $(u_i, v_i, d_i)$ of $V^i$, we first transform it to the novel view's voxel coordinate and tri-linearly sample the reconstructed novel view's density volume $V_{density}$:
\begin{equation}
    (u'_{novel}, v'_{novel}, d'_{novel})=C_{novel}(C_i^{-1}(u_i, v_i, \zeta(d_i))),
\end{equation}
\begin{equation}
    V_{density}^i(u_i, v_i, d_i) = V_{density}(u'_{novel}, v'_{novel}, d'_{novel}),
\end{equation}
where $(u_i, v_i, d_i)$ is the voxel coordinate in view $i$ and \\
$(u'_{novel}, v'_{novel}, d'_{novel})$ is the corresponding voxel coordinate in the novel view.
This volume re-sampling operation is shown in the first, second, and third columns of Figure~\ref{fig:visibility}.

After getting the density volume $V_{density}^i$ of input view $i$, we normalize it to $[0, 1]$ and get the alpha volume:
\begin{equation}
V_{alpha}^i(u_i, v_i, d_i) = 1-exp(-V_{density}^{i}(u_i, v_i, d_i)).  
\label{eq:alpha}
\end{equation}
Each voxel value of $V_{alpha}^i$ indicates the probability of surfaces existing in that voxel.
It also can be seen as the probability of a ray hitting a particle within that voxel.

Finally, for each sampled point $P_s \in N_h$, we transform it to each input view's voxel coordinate and tri-linearly sample the visibility volume to get the visibility weight:
\begin{equation}
(x_{i}^{'},y_{i}^{'},z_{i}^{'})=C_{i}(C_{novel}^{-1}(x,y,z)),
\end{equation}
\begin{equation}
    v^i=V_{vis}^{i}(x_{i}^{'},y_{i}^{'},\zeta^{-1}(z_{i}^{'})),
\end{equation}
\begin{equation}
    \zeta^{-1}(z_{i}^{'})=\frac{z_{i}^{'}-t_n}{t_f-t_n}*D,
\end{equation}
where $(x, y, z)$ is the coordinate of $P_s$ in novel view and $(x_{i}^{'},y_{i}^{'},z_{i}^{'})$ is the corresponding coordinate in view $i$.
The process of interpolating visibility weights $\{v_i\}_{i=1}^N$ is shown in the fifth and sixth columns of Figure~\ref{fig:visibility}.

In this way, we get the visibility weights ${\{v_i\}}_{i=1}^N$  to perform occlusion-aware multi-view feature aggregation in the texture volume, which helps to improve the rendering quality on the occlusion edges and areas.

\subsubsection{Ray integration}
Different from ENeRF \cite{enerf} or MVSNeRF \cite{mvsnerf}, which perform ray integration on color space, we conduct ray integration on feature space:
\begin{equation}
    f^T = \sum_{s=1}^{N_{h}} T_s(1-exp(-\sigma_s\delta_s))f_s^T,
\end{equation}
where $T_s=exp(-\sum_{j=1}^{s-1}\sigma_j\delta_j)$ is the transmittance value and $\delta_s$ is the distance between adjacent samples. Apart from a feature map, we also integrate the sampled and aggregated RGB values in Equation~\ref{eq:aggregation}:
\begin{equation}
    f^{RGB} = \sum_{s=1}^{N_{h}}T_s(1-exp(-\sigma_s\delta_s))f_s^{RGB}.
\end{equation}

After performing the ray integration for all the pixels in the novel view image, we get an interpolated color image $I_{novel}^{inter}\in\mathbb{R}^{H_T\times{W_T}\times3}$ and a feature map $F_{novel}^{lr}\in\mathbb{R}^{H_T\times{W_T}\times{C_T}}$, 
where $H_T, W_T$ are hyper-parameters and $C_T=16$ is the texture feature channels.

\begin{figure*}
    \centering
    \includegraphics[width=1.0\linewidth]{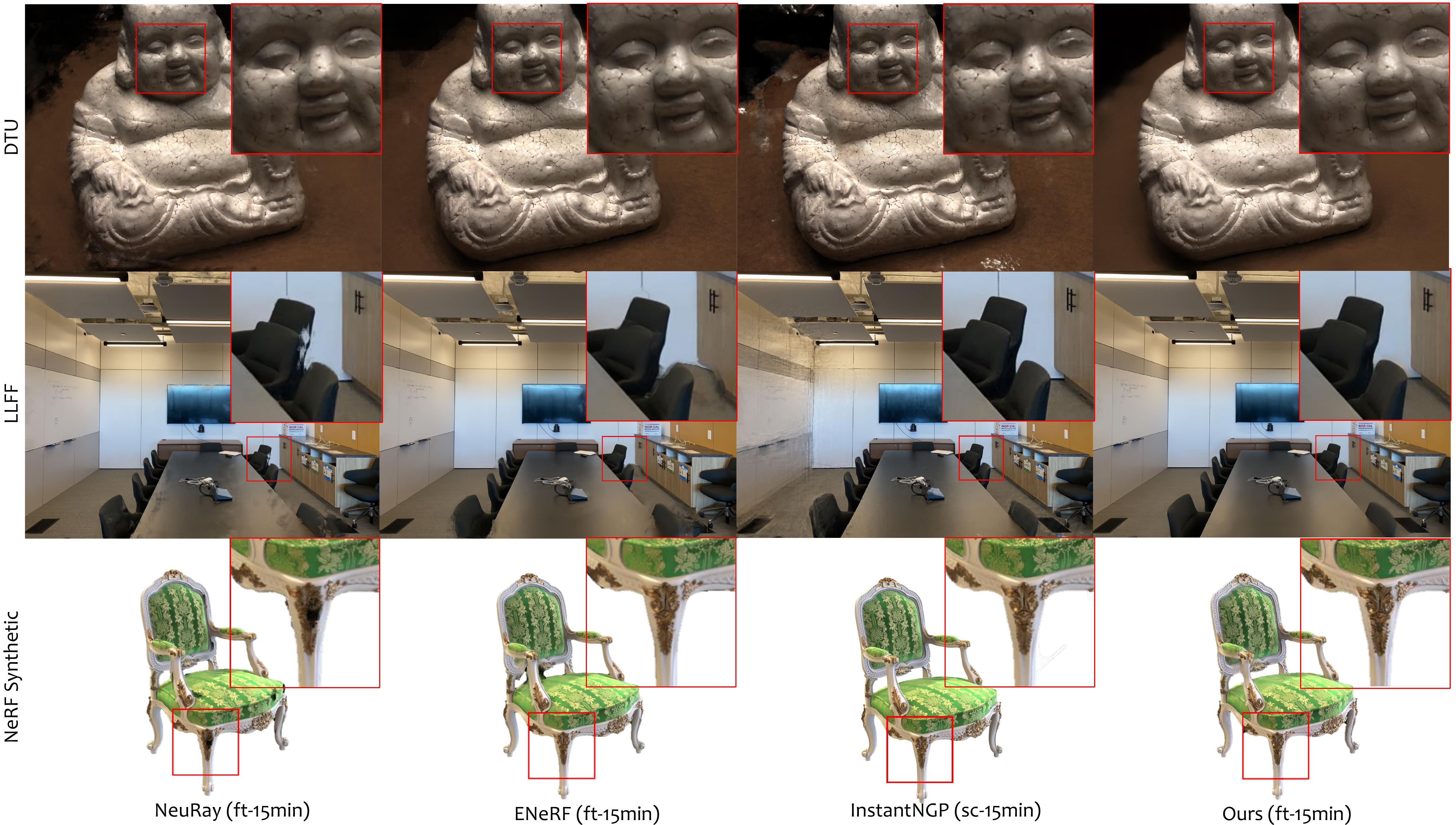} 
    \vspace{-0.4 cm}
    \caption{\textbf{Visual comparisons on static scenes.}  "ft-15min" means fine-tuning the generalization models for $\sim$15 minutes. 
    "sc-15min" means training networks from scratch for $\sim$15 minutes.
    Our method generally shows competitive rendering results with the baselines. 
    On the occluded areas of the scene in the LLFF dataset, our method achieves better results compared with previous generalizable methods (ENeRF, Neuray), demonstrating the efficiency of the proposed explicit 3D visibility reasoning.
    (Best viewed with zooming in on the page.)}
    \label{fig:static}
\end{figure*}

\begin{table*}[]
\caption{
\textbf{Quantitative evaluation on static datasets.} 
}
\begin{tabular}{c|l|ccc|ccc|ccc}
\toprule
\multirow{2}{*}{Settings}               & \multicolumn{1}{c|}{\multirow{2}{*}{Methods}} & \multicolumn{3}{c|}{DTU}                               & \multicolumn{3}{c|}{NeRF Synthetic}                    & \multicolumn{3}{c}{Real Forward-facing   LLFF}         \\ \cline{3-11} 
                                        & \multicolumn{1}{c|}{}                         & PSNR $\uparrow$ & SSIM $\uparrow$ & LPIPS $\downarrow$ & PSNR $\uparrow$ & SSIM $\uparrow$ & LPIPS $\downarrow$ & PSNR $\uparrow$ & SSIM $\uparrow$ & LPIPS $\downarrow$ \\ \hline
\multirow{4}{*}{Generalization}         & MVSNeRF                                       & 18.08           & 0.763           & 0.484              & 21.81           & 0.807           & 0.221              & 17.18           & 0.665           & 0.528              \\
                                        & NeuRay                                        & 20.11           & 0.848           & 0.371              & 24.32           & 0.892           & 0.152              & \textbf{21.30}  & \textbf{0.800}  & 0.335              \\
                                        & ENeRF                                         & 23.90           & \textbf{0.881}  & 0.291              & \textbf{25.81}  & \textbf{0.919}  & \textbf{0.106}     & 21.05           & 0.792           & \textbf{0.325}     \\
                                        & Ours                                          & \textbf{24.01}  & 0.875           & \textbf{0.289}     & 22.61           & 0.444           & 0.347              & 19.32           & 0.740           & 0.453              \\ \hline
\multirow{5}{*}{Per-scene optimization} & MVSNeRF$_{ft-15min}$                          & 19.04           & 0.768           & 0.470              & 24.19           & 0.890           & 0.153              & 21.55           & 0.776           & 0.397              \\
                                        & NeuRay$_{ft-15min}$                           & 21.17           & 0.856           & 0.365              & 24.63           & 0.890           & 0.145              & 21.51           & 0.805           & 0.325              \\
                                        & ENeRF$_{ft-15min}$                             & 24.20           & 0.887           & 0.286              & 26.27           & 0.924           & 0.110              & 22.40           & 0.803           & 0.304              \\
                                        & InstantNGP$_{sc-15min}$                       & \textbf{25.10}  & \textbf{0.907}  & 0.280              & \textbf{26.71}  & \textbf{0.933}  & 0.119              & 22.37           & 0.807           & 0.315              \\
                                        & Ours$_{ft-15min}$                             & 24.83           & 0.891           & \textbf{0.262}     & 26.54           & 0.927           & \textbf{0.086}     & \textbf{22.45}  & \textbf{0.812}  & \textbf{0.262}     \\ \bottomrule
\end{tabular}

\raggedright
"ft-15min" means fine-tuning pre-trained networks for nearly 15 minutes. 
"sc-15min" means learning networks from scratch for nearly 15 minutes.
Our method shows competitive rendering quality after fine-tuning for a short time.
\label{tab:static}
\end{table*}
\begin{figure}[]
    \centering
    \includegraphics[width=1\linewidth]{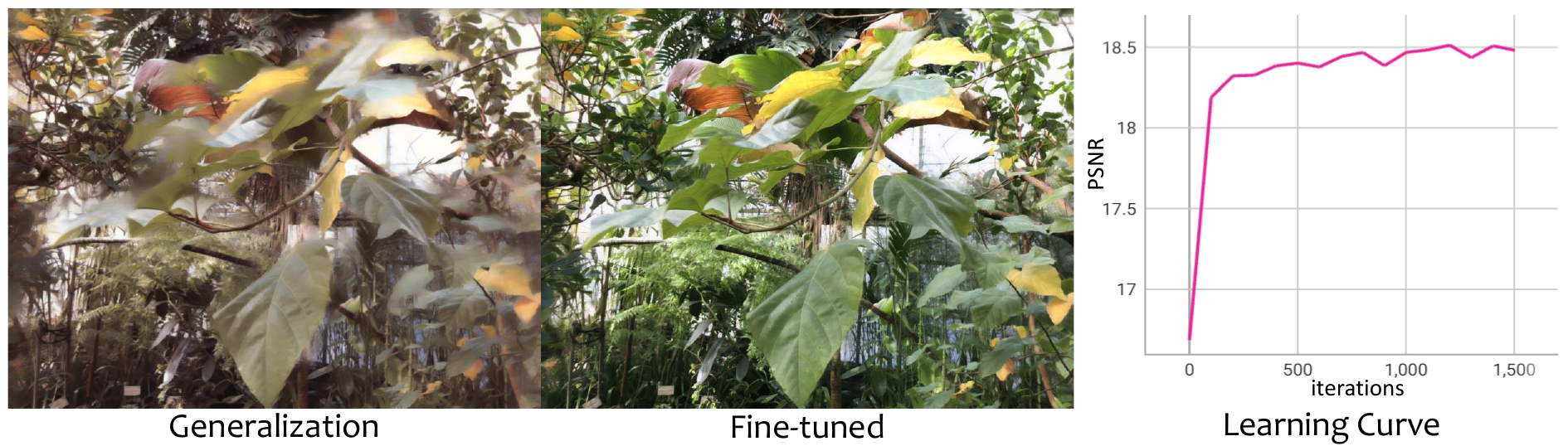}
    \vspace{-0.5 cm}
    \caption{\textbf{Fine-tuning on the "Leaves" scene.} 
    As the render-net $\mathcal{R}$ is a texture prior of the training data and the "Leave" scene in the LLFF dataset has very different texture from the training scenes in the DTU dataset, the rendering quality of generalization model decreases. 
    We show that the rendering quality can be significantly improved after fine-tuning for only 1500 iterations (\textasciitilde15 min).}
    \vspace{-0.2 cm}
    \label{fig:ft}
\end{figure}

\subsection{Rendering}
\label{sec:rendering}
After getting the novel-view feature map $F_{novel}^{lr}$, we use a 2D render network $\mathcal{R}$ to render the final image. To render high-resolution color images, we procedurally upsample $F_{novel}^{lr}$ to high-resolution in $\mathcal{R}$, as:
\begin{equation}
    I_{novel}^{hr} = \mathcal{R}(F_{novel}^{lr}).
    \label{eq:renderer}
\end{equation}
This rendering network $\mathcal{R}$ can serve as a texture prior of the rendered scenes, helping to improve the rendering quality and stability of high-resolution images.

\begin{figure*}[]
    \centering
    \includegraphics[width=\linewidth]{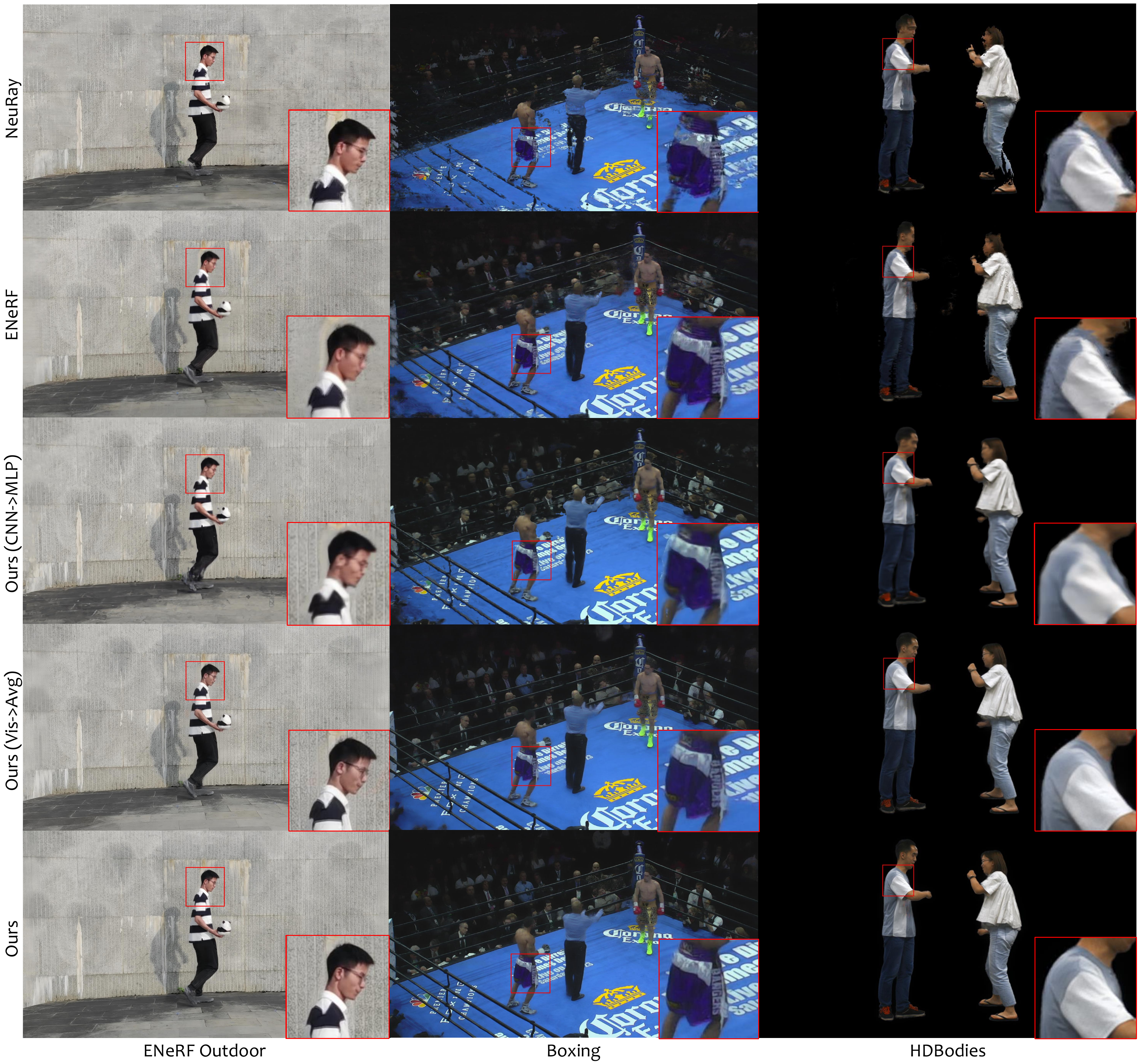} 
    \caption{\textbf{Visual comparisons on dynamic scenes.} 
    "CNN$\rightarrow$MLP" means replacing the render-net $\mathcal{R}$ with an MLP borrowed from vanilla NeRF.
    "Vis$\rightarrow$Avg" means replacing our explicit 3D visibility reasoning with average pooling operation.
    Benefiting from the explicit 3D visibility and the render-net $\mathcal{R}$, our method is able to render high-quality images on complex dynamic scenes, especially on the occluded edges and areas. (Best viewed with zooming in on the page.)} 
    \label{fig:dynamic}
\end{figure*}
\begin{figure}[ht]
    \centering
    \includegraphics[width=\linewidth]{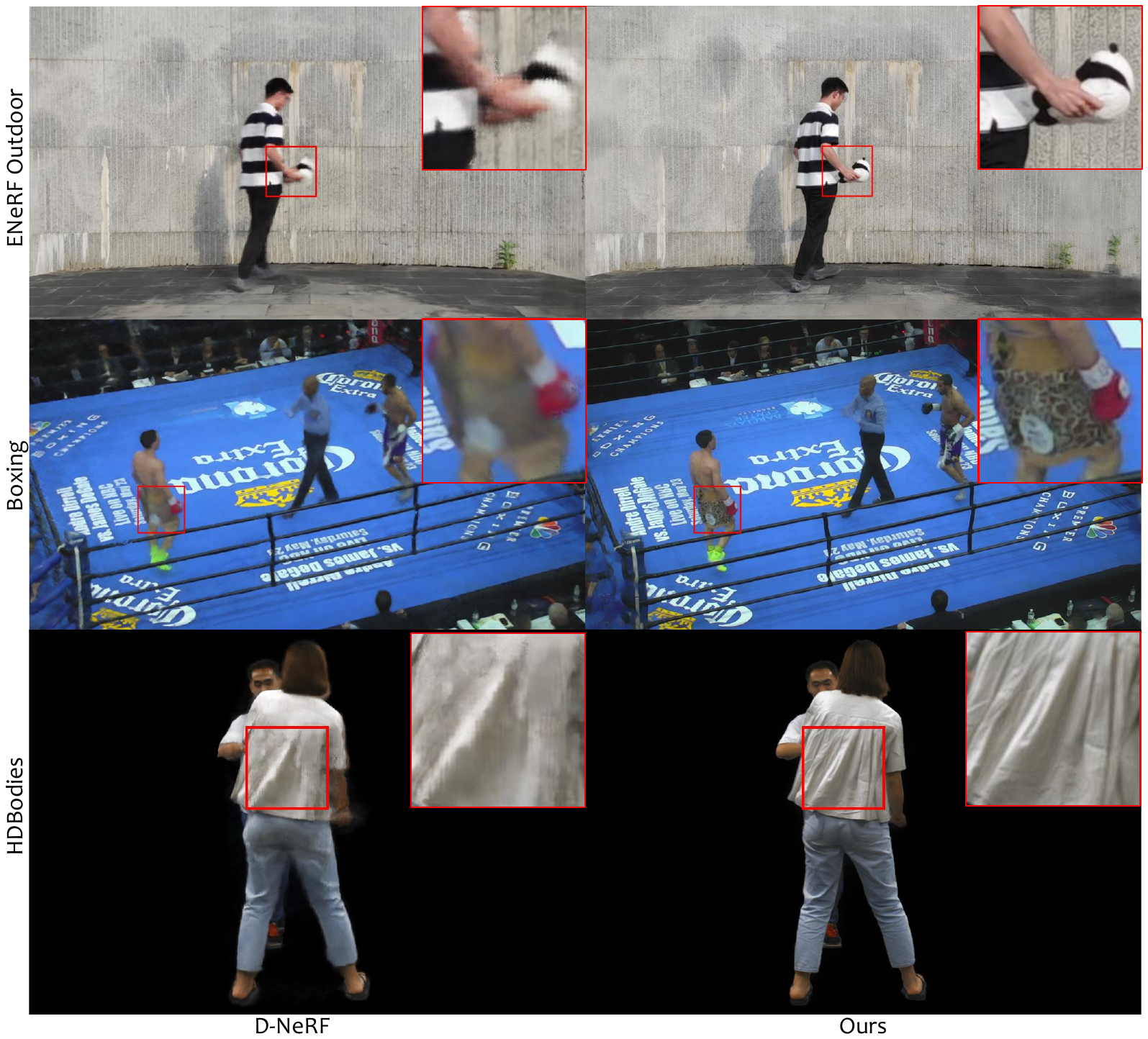} 
    \caption{\textbf{Visual comparisons with D-NeRF.}
    D-NeRF fails to recover the high-frequency texture details in the captured scenes whereas our method shows better results in rendering the fine-grained texture.}
    \label{fig:dnerf}
    \vspace{-0.2 cm}
\end{figure}

\subsection{Training Loss}
\label{sec:training_loss}
We train our model by minimizing the $\mathcal{L}_2$ loss and perceptual loss \cite{perceptualloss} between the rendered image $I_{novel}^{hr}$ and the ground-truth image $I_{gt}^{hr}$:
\begin{equation}
    \mathcal{L}_{render} = \Vert I_{novel}^{hr} - I_{gt}^{hr} \Vert_2^2,
\end{equation}
\begin{equation}
    \mathcal{L}_{percep} = \Vert \Phi(I_{novel}^{hr}) - \Phi(I_{gt}^{hr}) \Vert,
\end{equation}
where $\Phi$ is a pretrained VGG-19 network \cite{vgg19}, following \cite{svs}.

We also use an auxiliary loss inspired by \cite{lightfield} to make our model more geometrically interpretable. Specifically, we  we minimize the $\mathcal{L}_2$ loss between the interpolated image $I_{novel}^{inter}$ and  ground-truth $I_{gt}^{lr}$:
\begin{equation}
    \mathcal{L}_{inter} = \Vert I_{novel}^{inter} - I_{gt}^{lr} \Vert_2^2.
\end{equation}
This loss helps the geometry volumes to learn geometry from the input images. 

The total loss for training our network is:
\begin{equation}
    \mathcal{L}_{total} = \mathcal{L}_{render} + \mathcal{L}_{inter} + \lambda\mathcal{L}_{percep},
\end{equation}
where hyper-parameter $\lambda$ is set it to 0.1 in all the experiments.

\section{Experiments}
\subsection{Implementation Details}
Our implementation of the proposed method is based on the PyTorch \cite{pytorch} framework. Regarding the hyperparameters related to the resolution, specifically $H_G$ and $W_G$ in Section \ref{sec:geometry_volumes}, as well as $H_T$ and $W_T$ in Section \ref{sec:texture_volumes}, we have found that setting them to $H_G = H_{hr}/16$, $W_G = H_{hr}/16$, $H_T = H_{hr}/4$, and $W_T = H_{hr}/4$ strikes a good balance between rendering quality and speed in most cases. However, depending on the dataset being used, slight adjustments to these settings might be necessary to achieve optimal rendering results while maintaining real-time performance.

\subsection{Compared Methods and Evaluation Metrics}
To validate our method, we conduct experiments on both static and dynamic scenes for qualitative and quantitative evaluation.
As a generalizable rendering method, we mainly compared with recent state-of-the-art NeRF-based generalizable rendering methods, including MVSNeRF \cite{mvsnerf}, ENeRF \cite{enerf} and NeuRay \cite{neuralrays}.
MVSNeRF combines neural radiance fields with learning-based MVS to enable efficient radiance field construction.
ENeRF uses a learned depth-guided sampling strategy to speed up the generalizable radiance field methods.
NeuRay proposes an occlusion-aware method for constructing radiance fields.
Apart from generalizable rendering methods, we also compare with scene-specific and sequence-specific rendering methods to provide a more absolute calibration on the achievable quality of our method.
For scene-specific methods, we compare with InstantNGP \cite{instantngp} on static scenes, which uses multi-resolution hash encoding to accelerate training and rendering speed. 
Note that when dealing with dynamic scenes, though both generalizable rendering methods and scene-specific methods treat each frame as an independent scene, the scene-specific methods need to train a network for each frame individually whereas the generalizable rendering methods only have to train one network for each sequence as they can generalize across frames.
Training a network for each frame individually is impractical so we only perform comparisons with InstantNGP on static scenes.
For sequence-specific methods, we compare with D-NeRF \cite{dnerf} on dynamic scenes, which learns a deformation field for each sequence to map the sampled 3D points to a canonical space, then render images on the canonical field.
As a sequence-specific method, D-NeRF needs a sequence as training data to train a network, so we only perform comparisons with it on dynamic scenes.

For quantitative evaluation, we adopt the three widely used metrics: peak signal-to-noise ratio (PSNR), structural similarity index (SSIM) \cite{ssim} and learned perceptual image patch similarity (LPIPS) \cite{lpips}.
PSNR simply measures pixel-level accuracy, while SSIM and LPIPS reflect human perception better.

\begin{table}[t]
\caption{\textbf{Quantitative results on dynamic datasets.} } 
\resizebox{\linewidth}{!}{
\begin{tabular}{l|cc|cc|cc}
\toprule
\multirow{2}{*}{Methods}    & \multicolumn{2}{c|}{Ourdoor}         & \multicolumn{2}{c|}{Boxing}          & \multicolumn{2}{c}{HDBodies}         \\
                            & SSIM $\uparrow$ & LPIPS $\downarrow$ & SSIM $\uparrow$ & LPIPS $\downarrow$ & SSIM $\uparrow$ & LPIPS $\downarrow$ \\ \hline
MVSNeRF                     & 0.701           & 0.395              & 0.760           & 0.512              & 0.957           & 0.064              \\
NeuRay                      & 0.667           & 0.436              & 0.802           & 0.425              & 0.973           & 0.044              \\
ENeRF                       & 0.705           & 0.400              & 0.858           & 0.364              & 0.969           & 0.059              \\ \hline
Ours (CNN $\rightarrow$ MLP) & 0.725           & 0.443              & 0.863           & 0.381              & 0.973           & 0.049              \\
Ours (Vis $\rightarrow$ Avg) & 0.777           & 0.229              & 0.894           & 0.273              & 0.974           & 0.029              \\
Ours                        & \textbf{0.849}           & \textbf{0.185}              & \textbf{0.905}           & \textbf{0.245}              & \textbf{0.976}           & \textbf{0.028}              \\ \bottomrule
\end{tabular}}
\raggedright
CNN $\rightarrow$ MLP means replacing our render-net $\mathcal{R}$ with MLP. Vis $\rightarrow$ Avg means replacing our point-wise visibility reasoning with average pooling operation.
\label{tab:dynamic}
\end{table}

\subsection{Comparisons on Static Scenes} \label{sec:static}
\subsubsection{Datasets} For static scenes, we use three widely used benchmarks for evaluation: the DTU \cite{dtu}, NeRF synthetic \cite{llff}, and Real Forward-facing datasets \cite{nerf}.
We use the same train-test splits as in MVSNeRF and ENeRF. 
The resolution we use are 1600$\times$1200, 800$\times$800, and 1600$\times$1200 for the DTU, NeRF synthetic, and Real Forward-facing datasets respectively.

\subsubsection{Training} \label{sec:training_static}
For generalizable rendering methods, including MVSNeRF, ENeRF, NeuRay, and our method, we perform large-scale training on the DTU dataset and test the generalization models (only train on DTU) on all the three datasets to evaluate the generalization abilities.
We train our generalization model in an Nvidia Tesla A100 GPU with 40 GB GPU memory for 500K iterations with a batch size of 1.
For each iteration, we randomly pick one image as the novel view and three nearest neighbor images as the input views.
The training procedure costs nearly 1 day, using an Adam \cite{adam111} optimizer with a learning rate of $5\times10^{-4}$.
About the large-scale training of the alternative generalizable rendering methods, we use the default training settings in their implementation and the training procedures generally cost 1\textasciitilde2 days until their learning curves have converged.

Then, based on the pre-trained generalization models, we perform per-scene fine-tuning for a short time on each test scene of the three datasets.
Following MVSNeRF \cite{mvsnerf} and ENeRF \cite{enerf}, we fine-tune the generalization models to each test scene for \textasciitilde15 minutes.
In the fine-tuning stage, we optimize all the network parameters of our pipeline, including the encoder-net $\mathcal{E}$, the 3D CNN $\mathcal{O}$, and the render-net $\mathcal{R}$.
The training settings of the fine-tuning stage simply follow the settings of the large-scale training.

For InstantNGP, we train a network from scratch for \textasciitilde15 minutes on each test scene of the three datasets, using the default training settings in its implementation.

\subsubsection{Results}
The quantitative results of the generalization models are reported in the generalization setting rows of Table~\ref{tab:static}. 
The results show that under the generalization setting, our method shows competitive results on the DTU dataset, whereas on the LLFF and NeRF synthetic datasets, our method is inferior to previous methods. 
The reason for these results is that our render-net $\mathcal{R}$ can be regarded as a texture prior of the training set, and the test scenes in the DTU dataset have similar texture and illumination with the training scenes, so our render-net $\mathcal{R}$ is able to render high-quality results.
However, the new scenes in the NeRF synthetic and Real Forward-facing datasets have very different texture and illumination, so the rendering quality of our generalization model drops.

The quantitative results of the fine-tuned networks and the learn-from-scratch networks are reported in the per-scene optimization rows of  Table~\ref{tab:static}. 
The visual comparisons are illustrated in Figure~\ref{fig:static}.
These results demonstrate that after fine-tuning the generalization model to a new scene for a short time (\textasciitilde15 minutes), our method is able to achieve competitive results with previous works.
The fine-tuning procedure of the "Leaves" scene in the Real Forward-facing dataset is presented in Figure~\ref{fig:ft}, showing that the rendering quality of our method can be improved quickly during fine-tuning and after fine-tuning for only 1500 iterations (\textasciitilde15 minutes), the rendering quality is greatly enhanced.

\begin{table}[]
\caption{\textbf{Comparisons on rendering speed and memory consumption.} }
\resizebox{\linewidth}{!}{\begin{tabular}{l|cc|c|c}
\toprule
\multirow{2}{*}{Methods} & \multicolumn{2}{c|}{FPS $\uparrow$} & \multirow{2}{*}{Memory (MB)} & \multirow{2}{*}{\#Param.} \\ \cline{2-3}
                         & 720p        & 1080p      &                         &                          \\ \hline
MVSNeRF \cite{mvsnerf}                  & 0.05        & 0.02       & 20496.5                 & 0.13M                    \\
NeuRay \cite{neuralrays}                   & 0.03        & 0.02       & 7214.8                  & 4.75M                    \\
ENeRF \cite{enerf}                    & 5.93        & 2.84       & 9451.7                  & 0.44M                    \\
DNeRF \cite{dnerf}                    & 0.02        & 0.01       & 18555.7                 & 1.1M                     \\
InstantNGP \cite{instantngp}               & 55.17       & 27.65      & 1269.9                  & 11.5M                    \\
Ours                     & 27.08       & 16.12      & 2524.3                  & 3.01M                       \\ \bottomrule
\end{tabular}}

\raggedright
"720p" means rendering high-definition (HD) images at a resolution of $1280\times720$. "1080p" means rendering full high-definition (FHD) images at a resolution of $1920\times1080$. "Memory" means the GPU memory consumption when rendering 1080p images. "\#Param." means the number of network parameters. All the statics are tested on a single Nvidia Tesla A100 GPU.
\label{tab:fps}
\end{table}

\subsection{Comparisons on Dynamic Scenes}

\subsubsection{Datasets} 
\label{sec:dynamic_datasets}
For dynamic scenes, we perform comparisons on three datasets: the ENeRF Outdoor, Boxing, and HDbodies datasets. 
The ENeRF Outdoor dataset is an open-source dataset that is collected by ENeRF, which has 18 views in total and the average angular difference between the cameras is $6.66^\circ$. 
The Boxing dataset is proposed in \cite{su}, which uses 32 synchronized cameras in a circle arrangement to capture a boxing game, in which the captured scenes are highly complex and there are close human interactions between the two boxers. 
The HDbodies dataset is proposed in HDhuman \cite{hdhuman}, which uses 24 synchronized cameras in a circle arrangement to capture single or multi-human performers with complex texture patterns. 
The average angular difference between the input cameras is $11.25^\circ$ and $15^\circ$ for the Boxing and HDBodies datasets respectively. 
We use the first/last 900/100, 500/50, and 500/50 frames as the train/test frames for the ENeRF Outdoor, Boxing, and HDBodies datasets respectively.
The rendering resolution are $1920\times1080$, $1920\times1080$, and $1336\times1152$ for the ENeRF Outdoor, Boxing, and HDBodies datasets respectively.

\subsubsection{Training}
For generalizable rendering methods, we adopt the pre-trained generalization networks in Section~\ref{sec:training_static} as the initialization models, and fine-tune the networks to each dynamic scene. 
For each dynamic scene, we fine-tune our model for 50K iterations, which costs \textasciitilde3 hours. 
For each iteration, we randomly pick one frame in the training set, then pick one image as the novel view and four nearest neighbor images as the input views.
For fine-tuning the generalizable rendering baselines, we generally fine-tune them for 4\textasciitilde6 hours until the learning curves converge.

For D-NeRF, we found that training it on all frames of the training set of each sequence results in failures because there are too many frames in the training set of each sequence, overwhelming the deformation field of D-NeRF. 
Therefore, we select 100, 50, and 50 continuous frames on the ENeRF Outdoor, Boxing, and HDBodies datasets as the training data for D-NeRF.
Though we decrease the frames 
number of each sequence, the dynamic scenes in the three datasets are highly complex and contain complicated motions. This raises huge challenges for the deformation field of D-NeRF to represent the dynamic scenes, so it needs a long time to train the networks of D-NeRF. 
Concretely, for each dynamic scene, we train D-NeRF for 600K iterations, which costs \textasciitilde4 days.
When training D-NeRF, we follow the default training settings in their official implementation apart from the training iterations.
For fairness, to compare with D-NeRF, we fine-tune our generalization models in Section~\ref{sec:training_static} using the same training data of D-NeRF. 
Specifically, for each dynamic scene, we fine-tune our generalization model for 20K iterations, which costs \textasciitilde90 minutes.

\begin{table}[]
\caption{\textbf{The running time, memory consumption, and parameter number of different components of our pipeline.} }
\begin{tabular}{l|cc|c|c}
\toprule
\multirow{2}{*}{Components} & \multicolumn{2}{c|}{Time (ms)} & \multirow{2}{*}{Memory (MB)} & \multirow{2}{*}{\#Param.} \\ \cline{2-3}
                            & 720p           & 1080p         &                              &                           \\ \hline
Encoder-net $\mathcal{E}$   & $\sim$3.93     & $\sim$9.58    & 787.1                        & 0.18M                     \\
3D CNN $\mathcal{O}$        & $\sim$10.59    & $\sim$10.92   & 1069.6                       & 0.3M                      \\
Visibility reasoning        & $\sim$2.47     & $\sim$2.21    & -                            & -                         \\
Ray Integration             & $\sim$4.18     & $\sim$10.68   & -                            & -                         \\
Render-net $\mathcal{R}$    & $\sim$10.19    & $\sim$23.53   & 667.6                        & 2.53M                     \\ \hline
Total                       & $\sim$31.36    & $\sim$56.92   & 2524.3                       & 3.01M                     \\ \bottomrule
\end{tabular}

\raggedright
All the statics are tested on a single Nvidia Tesla A100 GPU.
The memory consumption is tested when rendering 1080p images. 
Note that the running time is not fully consistent with the rendering speeds that are reported in Table~\ref{tab:fps} as apart from the time consumption of the listed components, there are some additional consumption in our pipeline such as copying data.
\vspace{-0.2 cm}
\label{tab:components}
\end{table}

\subsubsection{Results} 
The quantitative and qualitative results of generalizable rendering methods are shown on the first three and last rows of Table~\ref{tab:dynamic} and Figure~\ref{fig:dynamic}.
Note that when rendering ENeRF Outdoor dataset, ENeRF uses an object-compositional representation for rendering the foreground and background independently, so that they can avoid the occlusion problems. 
However, the background is not always available for dynamic scenes, so we treat ENeRF Outdoor as a whole scene when training and testing our methods and all the baselines, without treating the foreground and background independently. So the rendering results are different from the results that are shown in ENeRF.
The results in Table~\ref{tab:dynamic} and Figure~\ref{fig:dynamic} demonstrate that our method outperforms previous works by a large margin when rendering complex dynamic scenes with sparse views, especially in the occluded edges and areas.
In the stage of multi-view feature aggregation, MVSNeRF naively calculates the variance of multi-view features without concerning any occlusion problems, making its results suffer from severe rendering artifacts and blurring with sparse views.
In ENeRF, it uses an MLP to predict visibility by utilizing the feature consistency between input views.
However, when the input views are relatively sparse and the scenes are highly complex, the multi-view features are highly inconsistent, making the results of ENeRF suffer from blurring in the occluded areas.
For NeuRay, its visibility reasoning procedure is actually performing per-view reconstruction and the reconstruction might be inconsistent between views.
When dealing with the highly complex Boxing dataset, the per-view reconstruction seems to be unreliable, making the inconsistency problems exacerbate.
As a result, the visibility estimation is inaccurate, so the rendering results of NeuRay in the Boxing dataset degrade dramatically.
Different from the implicit visibility reasoning methods of ENeRF and NeuRay, our explicit 3D visibility reasoning is globally consistent as all the visibility information is derived from the same novel view's density volume, providing accurate visibility estimation.
Therefore, our method shows the highest rendering quality, especially on the occluded edges and areas.

The visual comparisons with D-NeRF are illustrated in Figure~\ref{fig:dnerf}, showing that the rendered images of D-NeRF lose the high-frequency texture details whereas our method is able to recover the fine-grained texture.
This is because the dynamic scenes are in high resolution, containing complex texture and large motion, which makes the deformation field of D-NeRF difficult to map the sampled 3D points to the canonical space.
Therefore, D-NeRF fails to recover the high-frequency details of the rendered scenes, resulting in blurring in the rendered images. 
In contrast, our network can serve as the geometry and texture prior of the rendered scenes, which helps us to recover the fine-grained texture of the rendered scenes and improve the rendering quality.

\subsection{Running time and memory footprints analysis}
The comparisons of rendering speed, GPU memory consumption, and the number of network parameters are illustrated in Table~\ref{tab:fps}.
All the statistics are tested on a single Nvidia Tesla A100 GPU.
According to the statistics, InstantNGP achieves the fastest rendering speed while keeping a low memory consumption, but it can only handle static scenes.
MVSNeRF, NeuRay, and DNeRF run at extremely slow speeds as they need to query MLP millions of times.
ENeRF adopts a depth-guided sampling strategy to reduce the sampled points and as claimed in its paper, it can achieve real-time when rendering images at a resolution of $512\times512$.
However, ENeRF still needs to query the MLP for each sampled point once, so its rendering speed is inversely linear to the number of rendered pixels. 
When rendering high-resolution images, there are too many  pixels needed to render, which hinders it from real-time rendering.
In contrast, thanks to the ray integration on feature space that helps us bypass the time-consuming MLP query stage and the render-net $\mathcal{R}$ to generate high-resolution images, our method is able to render 720p images in real-time and render 1080p images at interactive frame rates. 
Besides, as shown in the supplementary video, our method is also able to render 1080p images at interactive frame rates using an Nvidia GeForce RTX 3090 GPU.
Regarding memory consumption, as we perform the intermediate volume processing in relatively low resolution, our method is able to render high-resolution images with a relatively low memory consumption.

We report the running time, memory footprints, and parameters number of each component of our pipeline in Table~\ref{tab:components}. Ideally, the time cost of rendering 1080p images should be twice of rendering 720p images.
However, we found that enlarging the geometry volume resolution of 1080p images has very slight improvements in rendering quality.
Therefore, we keep the resolution of geometry volumes such as density volumes and visibility volumes be same when rendering both 720p and 1080p images.
So the running time of the components such as 3D CNN and visibility reasoning that are related to the geometry volumes are nearly the same for both 720p and 
1080p images.
In contrast, for the components that belong to texture volumes, rendering 1080p images always costs more time than rendering 720p images.

\subsection{Ablation Studies}
\subsubsection{Effectiveness of the explicit 3D visibility reasoning} 
To evaluate the effectiveness of the proposed explicit 3D visibility reasoning, we replace it with average pooling operation and train on the three dynamic scenes. The quantitative and qualitative results are shown on the fifth row of Table~\ref{tab:dynamic} and Figure~\ref{fig:dynamic} respectively, demonstrating that the explicit 3D visibility reasoning helps to improve the rendering quality on the occluded edges or areas. 


\subsubsection{Effectiveness of the render-net $\mathcal{R}$}
To figure out the effectiveness of the render-net $\mathcal{R}$ in Section~\ref{sec:rendering}, we replace it with an MLP to render each pixel of the final image independently, in which the architecture of MLP is borrowed from vanilla NeRF. 
In the MLP-version pipeline, instead of integrating ray on feature space, we use MLP to decode each sampled point's aggregated feature to color and density values, and perform ray integration on the RGB space.
Without using a CNN to generate images, it is difficult to perform super-resolution operations, so we directly render images at high resolution and do not adopt any super-resolution technique in the MLP-version pipeline. 
We train the MLP-version pipeline on the three dynamic datasets and the quantitative and qualitative results are illustrated in the fourth row of Table~\ref{tab:dynamic} and Figure~\ref{tab:dynamic} respectively. The results show that our render-net $\mathcal{R}$ has better rendering quality as the 2D CNN can serve as a texture prior of the rendered scenes.


\section{Conclusions}
In this paper, we introduce a novel view synthesis method that achieves real-time rendering of high-resolution images.
At the heart of our method is explicit 3D visibility reasoning, which efficiently estimates the visibility of the sampled 3D points to input views, solving the occlusion problems arising from the sparsity of input views and the complexity of captured scenes.
Experimental results demonstrate that our visibility reasoning technique enhances the rendering quality, particularly in occluded edges and areas. 
Besides, with the carefully designed components in our pipeline such as the ray integration on feature space and the render-net $\mathcal{R}$ to serve a texture prior, our method achieves real-time speed in rendering high-resolution images.
Overall, our method enables the generation of high-quality novel-view images for complex dynamic scenes, even with a limited number of synchronized cameras from sparse views.

\section*{Acknowledgments}
This paper is supported by National Key R$\&$D Program of China (2022YFF0902200), the NSFC project to No.62125107.



 
%

\bibliographystyle{IEEEtran}
\bibliography{main}

\begin{thebibliography}{10}
\providecommand{\url}[1]{#1}
\csname url@samestyle\endcsname
\providecommand{\newblock}{\relax}
\providecommand{\bibinfo}[2]{#2}
\providecommand{\BIBentrySTDinterwordspacing}{\spaceskip=0pt\relax}
\providecommand{\BIBentryALTinterwordstretchfactor}{4}
\providecommand{\BIBentryALTinterwordspacing}{\spaceskip=\fontdimen2\font plus
\BIBentryALTinterwordstretchfactor\fontdimen3\font minus \fontdimen4\font\relax}
\providecommand{\BIBforeignlanguage}[2]{{%
\expandafter\ifx\csname l@#1\endcsname\relax
\typeout{** WARNING: IEEEtran.bst: No hyphenation pattern has been}%
\typeout{** loaded for the language `#1'. Using the pattern for}%
\typeout{** the default language instead.}%
\else
\language=\csname l@#1\endcsname
\fi
#2}}
\providecommand{\BIBdecl}{\relax}
\BIBdecl

\bibitem{nerf}
B.~Mildenhall, P.~P. Srinivasan, M.~Tancik, J.~T. Barron, R.~Ramamoorthi, and R.~Ng, ``{NeRF}: Representing scenes as neural radiance fields for view synthesis,'' in \emph{European Conference on Computer Vision}, 2020, pp. 405--421.

\bibitem{dnerf}
\BIBentryALTinterwordspacing
A.~Pumarola, E.~Corona, G.~Pons-Moll, and F.~Moreno-Noguer, ``{D-NeRF}: Neural radiance fields for dynamic scenes,'' in \emph{Proceedings of the IEEE/CVF Conference on Computer Vision and Pattern Recognition}, 2021. [Online]. Available: \url{http://arxiv.org/abs/2011.13961v1}
\BIBentrySTDinterwordspacing

\bibitem{hyperreel}
B.~Attal, J.-B. Huang, C.~Richardt, M.~Zollhoefer, J.~Kopf, M.~O’Toole, and C.~Kim, ``Hyperreel: High-fidelity 6-dof video with ray-conditioned sampling,'' in \emph{Proceedings of the IEEE/CVF Conference on Computer Vision and Pattern Recognition}, 2023, pp. 16\,610--16\,620.

\bibitem{gao2021dynamic}
C.~Gao, A.~Saraf, J.~Kopf, and J.-B. Huang, ``Dynamic view synthesis from dynamic monocular video,'' \emph{arXiv preprint arXiv:2105.06468}, 2021.

\bibitem{pixelnerf}
A.~Yu, V.~Ye, M.~Tancik, and A.~Kanazawa, ``{PixelNeRF}: Neural radiance fields from one or few images,'' in \emph{Proceedings of the IEEE/CVF Conference on Computer Vision and Pattern Recognition}, 2021, pp. 4578--4587.

\bibitem{ibrnet}
Q.~Wang, Z.~Wang, K.~Genova, P.~P. Srinivasan, H.~Zhou, J.~T. Barron, R.~Martin-Brualla, N.~Snavely, and T.~Funkhouser, ``{IBRNet}: Learning multi-view image-based rendering,'' in \emph{Proceedings of the IEEE/CVF Conference on Computer Vision and Pattern Recognition}, 2021, pp. 4690--4699.

\bibitem{enerf}
H.~Lin, S.~Peng, Z.~Xu, Y.~Yan, Q.~Shuai, H.~Bao, and X.~Zhou, ``Efficient neural radiance fields for interactive free-viewpoint video,'' in \emph{SIGGRAPH Asia 2022 Conference Papers}, 2022, pp. 1--9.

\bibitem{neuralrays}
Y.~Liu, S.~Peng, L.~Liu, Q.~Wang, P.~Wang, C.~Theobalt, X.~Zhou, and W.~Wang, ``Neural rays for occlusion-aware image-based rendering,'' in \emph{Proceedings of the IEEE/CVF Conference on Computer Vision and Pattern Recognition}, 2022, pp. 7824--7833.

\bibitem{mvsnerf}
A.~Chen, Z.~Xu, F.~Zhao, X.~Zhang, F.~Xiang, J.~Yu, and H.~Su, ``{MVSNeRF}: Fast generalizable radiance field reconstruction from multi-view stereo,'' in \emph{Proceedings of the IEEE International Conference on Computer Vision}, 2021.

\bibitem{fastnerf}
\BIBentryALTinterwordspacing
S.~J. Garbin, M.~Kowalski, M.~Johnson, J.~Shotton, and J.~Valentin, ``{FastNeRF}: High-fidelity neural rendering at 200fps,'' \emph{arXiv preprint arXiv:2103.10380}, 2021. [Online]. Available: \url{http://arxiv.org/abs/2103.10380v2}
\BIBentrySTDinterwordspacing

\bibitem{snerg}
\BIBentryALTinterwordspacing
P.~Hedman, P.~P. Srinivasan, B.~Mildenhall, J.~T. Barron, and P.~Debevec, ``Baking neural radiance fields for real-time view synthesis,'' \emph{arXiv preprint arXiv:2103.14645}, 2021. [Online]. Available: \url{http://arxiv.org/abs/2103.14645v1}
\BIBentrySTDinterwordspacing

\bibitem{mobilenerf}
Z.~Chen, T.~Funkhouser, P.~Hedman, and A.~Tagliasacchi, ``Mobilenerf: Exploiting the polygon rasterization pipeline for efficient neural field rendering on mobile architectures,'' in \emph{The Conference on Computer Vision and Pattern Recognition (CVPR)}, 2023.

\bibitem{instantngp}
T.~M{\"u}ller, A.~Evans, C.~Schied, and A.~Keller, ``Instant neural graphics primitives with a multiresolution hash encoding,'' \emph{arXiv preprint arXiv:2201.05989}, 2022.

\bibitem{dtu}
R.~Jensen, A.~Dahl, G.~Vogiatzis, E.~Tola, and H.~Aan{\ae}s, ``Large scale multi-view stereopsis evaluation,'' in \emph{Proceedings of the IEEE conference on computer vision and pattern recognition}, 2014, pp. 406--413.

\bibitem{llff}
B.~Mildenhall, P.~P. Srinivasan, R.~Ortiz-Cayon, N.~K. Kalantari, R.~Ramamoorthi, R.~Ng, and A.~Kar, ``Local light field fusion: Practical view synthesis with prescriptive sampling guidelines,'' \emph{ACM Transactions on Graphics (TOG)}, vol.~38, no.~4, pp. 1--14, 2019.

\bibitem{4knerf}
Z.~Wang, L.~Li, Z.~Shen, L.~Shen, and L.~Bo, ``4k-nerf: High fidelity neural radiance fields at ultra high resolutions,'' \emph{arXiv preprint arXiv:2212.04701}, 2022.

\bibitem{mipnerf}
\BIBentryALTinterwordspacing
J.~T. Barron, B.~Mildenhall, M.~Tancik, P.~Hedman, R.~Martin-Brualla, and P.~P. Srinivasan, ``Mip-nerf: A multiscale representation for anti-aliasing neural radiance fields,'' in \emph{Proceedings of the IEEE International Conference on Computer Vision}, 2021. [Online]. Available: \url{http://arxiv.org/abs/2103.13415v3}
\BIBentrySTDinterwordspacing

\bibitem{mipnerf360}
J.~T. Barron, B.~Mildenhall, D.~Verbin, P.~P. Srinivasan, and P.~Hedman, ``Mip-nerf 360: Unbounded anti-aliased neural radiance fields,'' in \emph{Proceedings of the IEEE/CVF Conference on Computer Vision and Pattern Recognition}, 2022, pp. 5470--5479.

\bibitem{dsnerf}
K.~Deng, A.~Liu, J.-Y. Zhu, and D.~Ramanan, ``Depth-supervised nerf: Fewer views and faster training for free,'' \emph{arXiv preprint arXiv:2107.02791}, 2021.

\bibitem{densedepth}
B.~Roessle, J.~T. Barron, B.~Mildenhall, P.~P. Srinivasan, and M.~Nie{\ss}ner, ``Dense depth priors for neural radiance fields from sparse input views,'' in \emph{Proceedings of the IEEE/CVF Conference on Computer Vision and Pattern Recognition}, 2022, pp. 12\,892--12\,901.

\bibitem{refnerf}
D.~Verbin, P.~Hedman, B.~Mildenhall, T.~Zickler, J.~T. Barron, and P.~P. Srinivasan, ``Ref-nerf: Structured view-dependent appearance for neural radiance fields supplemental material.''

\bibitem{pixelwise}
J.~L. Sch{\"o}nberger, E.~Zheng, J.-M. Frahm, and M.~Pollefeys, ``Pixelwise view selection for unstructured multi-view stereo,'' in \emph{Computer Vision--ECCV 2016: 14th European Conference, Amsterdam, The Netherlands, October 11-14, 2016, Proceedings, Part III 14}.\hskip 1em plus 0.5em minus 0.4em\relax Springer, 2016, pp. 501--518.

\bibitem{schoenberger2016sfm}
J.~L. Sch\"{o}nberger and J.-M. Frahm, ``Structure-from-motion revisited,'' in \emph{Conference on Computer Vision and Pattern Recognition}, 2016.

\bibitem{park2021nerfies}
\BIBentryALTinterwordspacing
K.~Park, U.~Sinha, J.~T. Barron, S.~Bouaziz, D.~B. Goldman, S.~M. Seitz, and R.~Martin-Brualla, ``Nerfies: Deformable neural radiance fields,'' in \emph{Proceedings of the IEEE International Conference on Computer Vision}, 2021. [Online]. Available: \url{http://arxiv.org/abs/2011.12948v5}
\BIBentrySTDinterwordspacing

\bibitem{nsvf}
\BIBentryALTinterwordspacing
L.~Liu, J.~Gu, K.~Z. Lin, T.-S. Chua, and C.~Theobalt, ``Neural sparse voxel fields,'' in \emph{Proceedings of the European Conference on Computer Vision (ECCV)}, 2020. [Online]. Available: \url{http://arxiv.org/abs/2007.11571v2}
\BIBentrySTDinterwordspacing

\bibitem{kilonerf}
\BIBentryALTinterwordspacing
C.~Reiser, S.~Peng, Y.~Liao, and A.~Geiger, ``Kilonerf: Speeding up neural radiance fields with thousands of tiny mlps,'' in \emph{Proceedings of the IEEE International Conference on Computer Vision}, 2021. [Online]. Available: \url{http://arxiv.org/abs/2103.13744v2}
\BIBentrySTDinterwordspacing

\bibitem{derf}
D.~Rebain, W.~Jiang, S.~Yazdani, K.~Li, K.~M. Yi, and A.~Tagliasacchi, ``Derf: Decomposed radiance fields,'' in \emph{Proceedings of the IEEE/CVF Conference on Computer Vision and Pattern Recognition}, 2021, pp. 14\,153--14\,161.

\bibitem{neuralsceneflow}
Z.~Li, S.~Niklaus, N.~Snavely, and O.~Wang, ``Neural scene flow fields for space-time view synthesis of dynamic scenes,'' in \emph{Proceedings of the IEEE/CVF Conference on Computer Vision and Pattern Recognition}, 2021, pp. 6498--6508.

\bibitem{du2021nerflow}
\BIBentryALTinterwordspacing
Y.~Du, Y.~Zhang, H.-X. Yu, J.~B. Tenenbaum, and J.~Wu, ``Neural radiance flow for {4D} view synthesis and video processing,'' in \emph{Proceedings of the IEEE International Conference on Computer Vision}, 2021. [Online]. Available: \url{http://arxiv.org/abs/2012.09790v2}
\BIBentrySTDinterwordspacing

\bibitem{viewbased}
K.~Pulli, H.~Hoppe, M.~Cohen, L.~Shapiro, T.~Duchamp, and W.~Stuetzle, ``View-based rendering: Visualizing real objects from scanned range and color data,'' in \emph{Rendering Techniques’ 97: Proceedings of the Eurographics Workshop in St. Etienne, France, June 16--18, 1997 8}.\hskip 1em plus 0.5em minus 0.4em\relax Springer, 1997, pp. 23--34.

\bibitem{parallax}
K.~C. Zheng, A.~Colburn, A.~Agarwala, M.~Agrawala, D.~Salesin, B.~Curless, and M.~F. Cohen, ``Parallax photography: creating 3d cinematic effects from stills,'' in \emph{Proceedings of Graphics Interface 2009}, 2009, pp. 111--118.

\bibitem{highquality}
C.~L. Zitnick, S.~B. Kang, M.~Uyttendaele, S.~Winder, and R.~Szeliski, ``High-quality video view interpolation using a layered representation,'' \emph{ACM transactions on graphics (TOG)}, vol.~23, no.~3, pp. 600--608, 2004.

\bibitem{soft3d}
E.~Penner and L.~Zhang, ``Soft {3D} reconstruction for view synthesis,'' \emph{ACM Transactions on Graphics (TOG)}, vol.~36, no.~6, pp. 1--11, 2017.

\bibitem{deepblending}
P.~Hedman, J.~Philip, T.~Price, J.-M. Frahm, G.~Drettakis, and G.~Brostow, ``Deep blending for free-viewpoint image-based rendering,'' \emph{ACM Transactions on Graphics (TOG)}, vol.~37, no.~6, pp. 1--15, 2018.

\bibitem{fvs}
G.~Riegler and V.~Koltun, ``Free view synthesis,'' in \emph{European Conference on Computer Vision}, 2020, pp. 623--640.

\bibitem{drebin1988volume}
R.~A. Drebin, L.~Carpenter, and P.~Hanrahan, ``Volume rendering,'' \emph{ACM Siggraph Computer Graphics}, vol.~22, no.~4, pp. 65--74, 1988.

\bibitem{perceptualloss}
J.~Johnson, A.~Alahi, and L.~Fei-Fei, ``Perceptual losses for real-time style transfer and super-resolution,'' in \emph{European Conference on Computer Vision}.\hskip 1em plus 0.5em minus 0.4em\relax Springer, 2016, pp. 694--711.

\bibitem{vgg19}
K.~Simonyan and A.~Zisserman, ``Very deep convolutional networks for large-scale image recognition,'' \emph{arXiv preprint arXiv:1409.1556}, 2014.

\bibitem{svs}
G.~Riegler and V.~Koltun, ``Stable view synthesis,'' in \emph{Proceedings of the IEEE Conference on Computer Vision and Pattern Recognition}, 2021.

\bibitem{lightfield}
M.~Suhail, C.~Esteves, L.~Sigal, and A.~Makadia, ``Light field neural rendering,'' in \emph{Proceedings of the IEEE/CVF Conference on Computer Vision and Pattern Recognition}, 2022, pp. 8269--8279.

\bibitem{pytorch}
A.~Paszke, S.~Gross, F.~Massa, A.~Lerer, J.~Bradbury, G.~Chanan, T.~Killeen, Z.~Lin, N.~Gimelshein, L.~Antiga \emph{et~al.}, ``Pytorch: An imperative style, high-performance deep learning library,'' \emph{Advances in neural information processing systems}, vol.~32, 2019.

\bibitem{ssim}
Z.~Wang, A.~C. Bovik, H.~R. Sheikh, and E.~P. Simoncelli, ``Image quality assessment: from error visibility to structural similarity,'' \emph{IEEE Transactions on Image Processing}, vol.~13, no.~4, pp. 600--612, 2004.

\bibitem{lpips}
R.~Zhang, P.~Isola, A.~A. Efros, E.~Shechtman, and O.~Wang, ``The unreasonable effectiveness of deep features as a perceptual metric,'' in \emph{Proceedings of the IEEE Conference on Computer Vision and Pattern Recognition}, 2018, pp. 586--595.

\bibitem{adam111}
D.~P. Kingma and J.~Ba, ``Adam: A method for stochastic optimization,'' \emph{arXiv preprint arXiv:1412.6980}, 2014.

\bibitem{su}
Z.~Su, T.~Zhou, K.~Li, D.~Brady, and Y.~Liu, ``View synthesis from multi-view rgb data using multilayered representation and volumetric estimation,'' \emph{Virtual Reality \& Intelligent Hardware}, vol.~2, no.~1, pp. 43--55, 2020.

\bibitem{hdhuman}
T.~Zhou, J.~Huang, T.~Yu, R.~Shao, and K.~Li, ``Hdhuman: High-quality human novel-view rendering from sparse views,'' \emph{IEEE Transactions on Visualization and Computer Graphics}, 2023.

\end{thebibliography}

\end{document}